\documentclass[global,twocolumn]{svjour}
\usepackage{graphicx}
\usepackage{siunitx}
\usepackage{braket}
\usepackage{lineno}
\usepackage{ulem} 

\begin{document}
\title{A compact and robust diode laser system for atom interferometry on a sounding rocket}
\author{V. Schkolnik\inst{1,5} \and O. Hellmig\inst{3} \and A. Wenzlawski\inst{2} \and J. Grosse\inst{4,6} \and A. Kohfeldt\inst{5} \and K. D\"oringshoff \inst{1}  \and A. Wicht\inst{1,5} \and P. Windpassinger\inst{2} \and K. Sengstock\inst{3}  \and C. Braxmaier\inst{4,6} \and M. Krutzik\inst{1} \and A. Peters\inst{1,5} 
}                     
%
%
\institute{ Institut f\"{u}r Physik, Humboldt-Universit\"{a}t zu Berlin, Newtonstr. 15, 12489 Berlin, Germany\\ \email{vladimir.schkolnik@physik.hu-berlin.de} 
\and Institut f\"{u}r Physik, Johannes Gutenberg-Universit\"{a}t Mainz, 55099 Mainz, Germany 
\and Institut f\"{u}r Laserphysik, Universit\"{a}t Hamburg, Luruper Chaussee 149, 22761 Hamburg, Germany 
\and Zentrum f\"{u}r angewandte Raumfahrttechnologie und Mikrogravitation (ZARM), Universit\"{a}t Bremen, Am Fallturm, 28359 Bremen,
 Germany
\and Ferdinand-Braun-Institut, Leibniz-Institut f\"{u}r H\"{o}chstfrequenztechnik,
Gustav-Kirchhoff-Str. 4, 12489 Berlin, Germany
\and Deutsches Zentrum f\"{u}r Luft- und Raumfahrt e.V. (DLR) Institut f\"{u}r Raumfahrtsysteme, Robert-Hooke-Str. 7, 28359 Bremen}
\date{}
%
\maketitle
\begin{abstract}

We present a diode laser system optimized for laser cooling and atom interferometry with ultra-cold rubidium atoms aboard sounding rockets as an important milestone towards space-borne quantum sensors. Design, assembly and qualification of the system, combing micro-integrated distributed feedback (DFB) diode laser modules and free space optical bench technology is presented in the context of the MAIUS (Matter-wave Interferometry in Microgravity) mission.

This laser system, with a volume of 21 liters and total mass of 27 kg, passed all qualification tests for operation on sounding rockets and is currently used in the integrated MAIUS flight system producing Bose-Einstein condensates and performing atom interferometry based on Bragg diffraction. The MAIUS payload is being prepared for launch in fall 2016.

We further report on a reference laser system, comprising a rubidium stabilized DFB laser, which was operated successfully on the TEXUS 51 mission in April 2015. The system demonstrated a high level of technological maturity by remaining frequency stabilized throughout the mission including the rocket's boost phase.

\end{abstract}

\section{Introduction}
\label{sec:intro}

Since their first realization in 1991 \cite{Kasevich1991a} light pulse atom interferometers have matured into usable quantum sensors such as gravimeters \cite{Peters2001}, gradiometers \cite{McGuirk2002} or gyroscopes \cite{Gustavson2000}. Along with technological developments atom interferometers have become mobile sensors with applications in geodesy \cite{Freier2015,Merlet2010,Gillot2014} and future developments will enable their utilization in the field for geophysics, seismology or navigation \cite{Fang2016}. Moreover, systems featuring atom interferometry with various atomic species have been developed for high precision tests of fundamental physics such as the Universality of Free Fall (UFF) as part of the Einstein equivalence principle (EEP) with single \cite{Peters1999,Muller2010} and dual species atom interferometers \cite{Schlippert2014,Zhou2015}.

As the sensitivity of an atom interferometer scales with the free evolution time, systems aiming at high precision use large drop towers or atomic fountains to achieve the necessary measurement time \cite{Muentinga2013,Dickerson2013,Hartwig2015}. Space missions ultimately allow for unprecedented interrogation times without the need for large system dimensions. Currently several international projects and studies are aiming for the realization of space based quantum sensors with potential applications in precision tests of fundamental physics \cite{Aguilera2014}, gravitational wave astronomy \cite{Graham2013} and gravity gradiometry based geodesy \cite{Yu2006,Carraz2014}. Ongoing projects prepare for atom interferometry based quantum sensors on board of the International Space Station or a satellite \cite{Aguilera2014,Hogan2011}. To perform these types of experiments, the payload's subsystems including the laser system need to be compact, robust and maintenance-free over the full time of the mission. Such laser systems are already in operation on earth-based microgravity platforms such as the Bremen drop tower or the Novespace zero-g airplane \cite{Muentinga2013,Geiger2011}. 
The latter laser system for atom interferometry with rubidium and potassium at \SI{780}{nm} and \SI{767}{nm} is based on lasers in the telecommunication band and passed several qualification steps for space applications \cite{Theron2014,Leveque2014}. Despite the high maturity of the telecommunication technology it requires a frequency doubling stage when used for atom interferometry with rubidium and potassium and the laser light switching and distribution has still to be performed in the near infrared region. Using diode lasers and optical amplifiers operating directly at the desired wavelengths is an alternative method without the need for frequency conversion modules. A laser system for laser cooling of cesium based on diode lasers has already been developed for operation of the Pharao atomic clock in the ACES mission \cite{Leveque2015}. 

Within the scope of several projects, funded by the German Aerospace Cente (DLR), all critical technologies for diode laser based systems to perform atom interferometry on a sounding rocket were developed. 

This paper is structured as follows: Section 2 introduces the sounding rocket environment and the resulting requirements for laser systems to be operated on one. In Section 3 we present the key components of our laser system. It will perform efficient laser cooling in a 2D+/3D magneto-optical trap (MOT) assembly, state preparation, atom interferometry based on Bragg pulses with $^{87}$Rb and detection within the MAIUS mission \cite{Grosse2014} during approximately six minutes of microgravity. Section 4 reports on the conducted qualification steps for the MAIUS laser system and on the optical reference on-board the TEXUS 51 sounding rocket to mission qualify the principal technologies before the MAIUS launch.

Measurements and performance details of the complete MAIUS payload will be presented in detail in future publications.

\section{Sounding rocket as a laboratory in space}
\label{sec:rocket}

The MAIUS mission has the goal to perform atom interferometry with Bose-Einstein condensates for the first time in a space environment. This will be done on a sounding rocket flight from Esrange Space Center in Sweden scheduled for fall 2016. The experiment will be launched on a two stage VSB-30 rocket \cite{dlr74326}. With the given mass of the experimental payload this rocket type will reach an apogee of \SI{238}{km} and provide a microgravity time of approximately 6 minutes \cite{Grosse2014}.

\begin{figure}[h]
\centering
\includegraphics[width=0.95\linewidth]{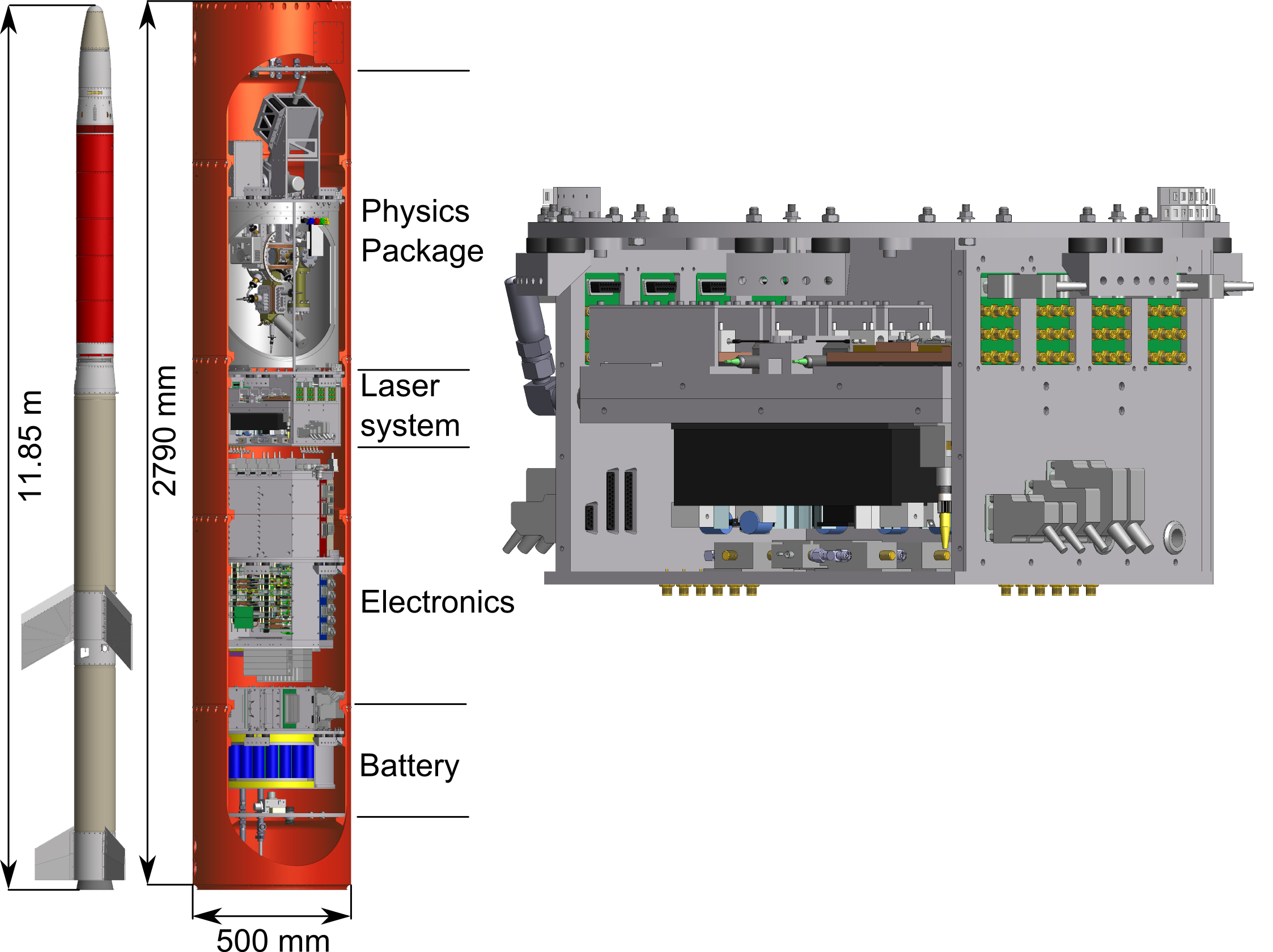}
\caption{CAD drawing of the MAIUS payload \cite{Stamminger2015} and the MAIUS laser system. The diameter of the hull sections is \SI{500}{mm}. }
\label{fig:spacelab}
\end{figure}

The whole payload, depicted in Figure \ref{fig:spacelab} (left), consists of a UHV chamber including an atom chip similar to the ones used for drop tower experiments \cite{Rudolph2015} and fiber optical access for 2D+/3D MOT, detection and Bragg interferometry beams. It is housed in five stacked hull segments with a diameter of \SI{500}{mm} together with the laser system, power supply and electronics. The laser system depicted in Figure \ref{fig:spacelab} (right) is controlled by a dedicated electronic system and provides light for initial cooling of $^{87}$Rb in a 2D+ MOT which loads a 3D MOT. The atoms in the 3D MOT are cooled further down using optical molasses are optically pumped for loading into the atom chip trap. After the evaporation to a Bose-Einstein condensate, atom interferometry is performed by a sequence composed of Bragg pulses \cite{Muntinga2013} and the population in the interferometer output ports is detected by absorption imaging.

The advantage of long free evolution times compared to usual laboratory environments comes at the price of very limited space for all subsystems, constraints on the overall mass and the lack of a stable temperature environment. It additionally requires high stability due to high mechanical stress on the payload during launch and re-entry. In spite of these conditions, all systems have to perform accurately during the limited time of the sounding rocket mission that starts after the boost phase of the rocket. This is particular challenging for the laser system due to the high mechanical loads. During the boost phase of approximately \SI{44}{s} the motors will cause accelerations of up to \SI{13}{g} and vibrational loads of around \SI{1.8}{g_{RMS}}. Therefore the system and its components have to be qualified in vibration tests at levels of \SI{5.4}{g_{RMS}} and \SI{8.1}{g_{RMS}}, respectively, as a requirement from the launch provider.

The laser system needs to be reliably operating and must not show a critical degradation after integration in the rocket hull until launch and finally during the complete flight. There is no physical access to the individual parts of the laser system during ground operation after the final assembly. Therefore it is required to remotely monitor all needed system parameters, especially the power levels in the individual optical paths.
Water cooling is available only before lift-off. After launch the temperature of the heat sink is controlled to about \SI{1}{K}. 
During the sounding rocket flight the temperature of the payload structure will increase by approximate \SI{5}{\degree C} and the thermal design of the laser system should guarantee operation at every period of the mission.

\section{Laser subsystem technology}
\label{sec:subsystem}

The previous section introduced the requirements for a laser system for rubidium BEC atom interferometry in space. In this section we discuss the key subsystems and describe the underlying technological approach.

\subsection{Micro-integrated diode laser modules}

The laser modules used for the MAIUS rocket mission are based on a versatile platform comprising semiconductor diode lasers and amplifiers, micro-optics and electronic interfaces altogether assembled onto robust micro-integrated optical benches (MIOB) \cite{Luvsandamdin2014} with a footprint of only \SI{80}{mm} $\times$ \SI{25}{mm}. These modules are explicitly designed for rubidium BEC and atom interferometry experiments on a sounding rocket. This platform can be transferred to other wavelengths of interest by using appropriate semiconductor chips. The modules feature a
master oscillator power amplifier (MOPA) configuration consisting of a distributed feedback diode laser (DFB) as the master oscillator (MO) with a linewidth of about \SI{1}{MHz} followed by a tapered amplifier (PA) generating an output power of more than \SI{1}{W}. The optical power emitted through the back facet of the DFB diode is monitored with a photo diode. Micro-isolators at the main output of the master oscillator with \SI{60}{dB} isolation and at the rear with \SI{30}{dB} isolation protect the DFB diode laser from optical feed-back. The DFB diode lasers operate at a temperature around \SI{36}{\degree C} and with injection currents of \SI{200}{mA}. The PA sections with the corresponding ridge-waveguide (RW) sections operate at \SI{2000}{mA} and \SI{200}{mA}, respectively.

\begin{figure}[h]
\centering
\includegraphics[width=0.99\linewidth]{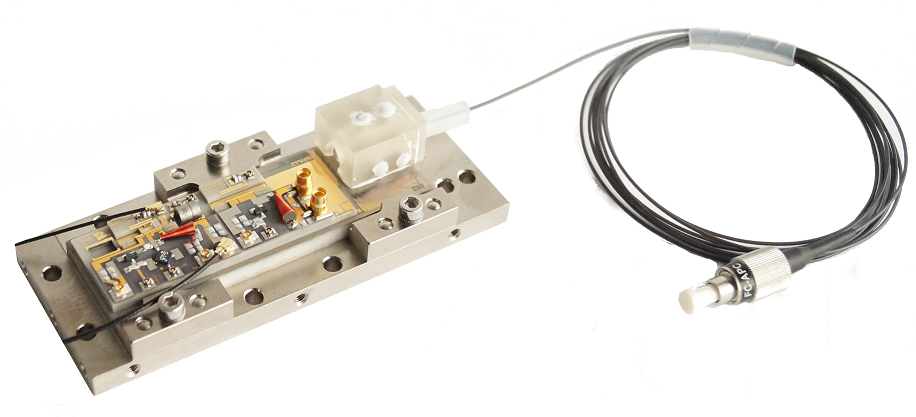}
\caption{The diode laser module used in the MAIUS laser system. A DFB diode is amplified by a tapered amplifier up to \SI{1}{W}. The output beam is coupled into a polarization maintaining fiber. Module footprint with the CCP: \SI{104}{mm} $\times$ \SI{42}{mm} (The PA is not integrated in the shown module).}
\label{fig:MOPA}
\end{figure}

The output of the PA is coupled into a polarization maintaining single mode fiber by means of a fiber coupler made from Zerodur as described in \cite{Duncker2014}. 
 To ensure high mechanical and thermal stability the MIOB is clamped on a conductively-cooled package (CCP) made from copper, which is mounted on a Peltier element for stabilization of the MIOB temperature. Figure \ref{fig:MOPA} shows a fiber coupled MOPA module mounted on such a CCP.
An extensive description of the employed laser technology can be found in \cite{Schiemangk2015}.

For operation of the MAIUS experiment four laser modules are used (see Figure \ref{fig:LS_SW}a). One of them is a DFB MO laser without a power amplifier, which is frequency stabilized to an atomic transition of rubidium and serves as the frequency reference for the three DFB-based MOPA modules (from now on called science lasers) as shown in Figure \ref{fig:LS_SW}b. The laser system contains two additional MOPA modules, electrically connected with a separate set of calibrated electronics. This modules can be used as a replacement for a science laser by a single splice connection in case of malfunction before the launch.

\subsection{Frequency stabilization}
\label{sec:stabil}

The light from every laser module passes a fiber-coupled optical isolator to reduce optical feedback and its power is monitored consecutively by means of an in-line photo diode. The light is then guided to polarization maintaining fiber splitters with a splitting ratio of 1:99 for the science lasers and 10:90 for the reference laser. The main part of the light from the reference laser is split into four usable ports, three of which are overlapped with the light from the science lasers (see Figure \ref{fig:LS_SW}c) in an all-fiber based splitter system. The smaller fraction of the light from the reference laser is guided to a frequency modulation spectroscopy (FMS) module based on Zerodur optical bench technology \cite{Duncker2014}. The spectroscopy module is used for stabilizing the reference laser to the $\ket{F = 3} \rightarrow \ket{F' = 3/4}$ crossover transition of $^{85}$Rb using  Doppler-free frequency modulation spectroscopy (FMS). An additional Doppler-broadened signal is generated as well and can be used for monitoring purposes.

The main part of the light from the science laser modules is guided to the switching and distribution module. The generated beat-notes are used for frequency offset stabilization of the science lasers. Light from the three overlapped output ports is guided onto fast photo diodes adhesively bonded to fiber collimators which are attached to the outputs of the beat splitter module.

\subsection{Switching and distribution}
\label{sec:switch}

The distribution of the light to the different fibers connected to the vacuum chamber as well as intensity switching and pulse shaping is realized in a switching and distribution module comprising a Zerodur-based free space switching module and a fiber-based distribution unit. The switching module is described in \cite{Duncker2014}. The module is realized on a Zerodur bench (with a footprint of \SI{210}{mm} $\times$ \SI{180}{mm}) with Zerodur based fiber collimators to inject the light into the module, HR- and AR-coated optics, acousto-optical modulators and finally Zerodur based fiber couplers at the module's output. For total extinction of spurious stray light a stepper motor driven shutter is implemented in front of every fiber coupler.

\begin{figure*}[htb]
\centering
\includegraphics[width=0.95\textwidth]{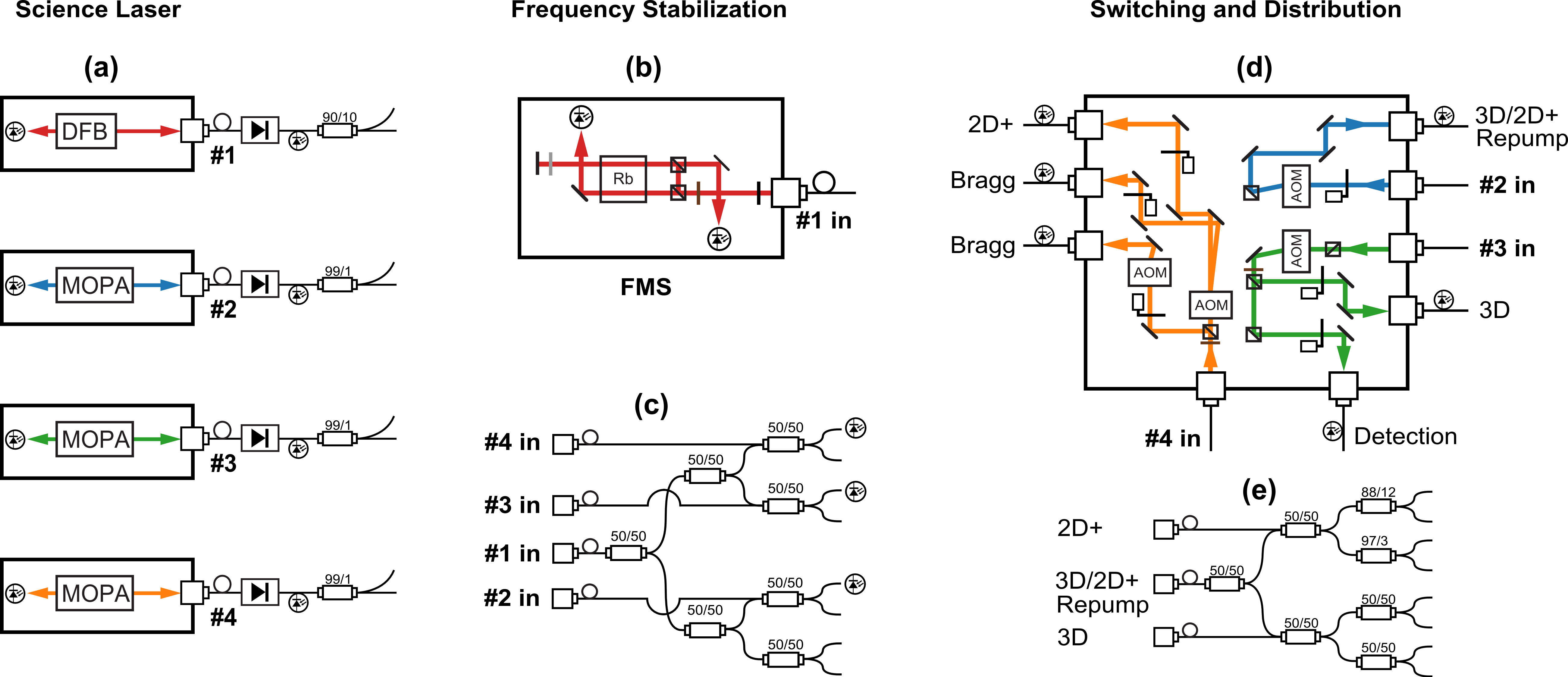}
\caption{The MAIUS laser system. The reference laser (\#1) is stabilized using FM spectroscopy (b) while the three science laser (\#2-\#4) are offset locked to the reference. The beat signals are generated by overlapping a small fraction of each laser light with the reference laser in the fiber based beat splitter system (c). The light generated by the three science lasers is distributed on an optical switching module bench made from Zerodur (d) with a footprint of \SI{210}{mm} $\times$ \SI{180}{mm}. The power splitting and overlapping for the 2D+ and 3D MOT is performed in a fiber splitter connected to the output of the Zerodur bench (e).}
\label{fig:LS_SW}
\end{figure*}

The module comprises the inputs from the three science lasers and six outputs connected with polarization maintaining fibers (see Figure \ref{fig:LS_SW}d). 
Light from laser \#2 passes an AOM and the first diffraction order is used as re-pumping light in the 2D+ and 3D MOT.
The light from laser \#3 passes an AOM and the first diffraction order is split into two paths. The first fiber output is used for 3D MOT cooling while the second is used for detection. The ratio between 3D cooling and detection light is adjusted by a wave plate and polarizing beam splitter (PBS) and remains fixed for the mission. Light from laser \#4 is utilized for the 2D+ MOT cooling as well as for the Bragg interferometry pulses. The cooling light for the 2D+ MOT passes a PBS and an AOM, where the zeroth diffraction order is used. Since there is no need for fast switching for the 2D+ MOT light, switching is realized using a shutter only. The Bragg pulses are generated by two AOMs driven around \SI{80}{MHz} with a frequency difference of n$\cdot$\SI{15}{kHz} corresponding to a multiple of n of the two photon recoil frequency \cite{Muentinga2013,Altin2013}. All AOMs except the Bragg ones are usually in the on-state while the shutters block the beams. Before a pulse, the AOMs are then switched off, the shutters are opened, a light pulse is generated with the AOM and the shutters are closed again. This way the AOMs remain at almost constant temperature. The fiber coupling for Bragg beams was maximized for pulsed operation.

The fiber coupled optical power in each output is monitored with an in-line photo diode. This way we gain information about the potential intrinsic losses of each subsystem throughout the mission and during the ground operation. The detection and both Bragg fibers are directly guided to the experiment,
whereas the 2D+ cooling, 3D cooling and 2D+/3D repumping light are overlapped in a 3\,to\,8 fiber based splitter system (see Figure \ref{fig:LS_SW}e). The splitter system is designed to overlap the repumping and cooling light with the necessary intensity ratios for the 2D+ and 3D MOT operation. The four beams for the 2D+ MOT have a ratio of 44\%, 48.5\%, 6\% and 1.5\% for the two transverse cooling beams, one pushing and one retardation beam, respectively, based on the design of a drop tower apparatus described in \cite{Rudolph2015}. The polarization extinction ratio (PER) per individual fiber coupler is better than \SI{-23}{dB}.  Four beams  with a symmetric splitting ratio of 25\% each are used for the 3D chip-MOT \cite{Hansel2001}. 

\begin{figure}[hb]
\centering
\includegraphics[width=0.95\linewidth]{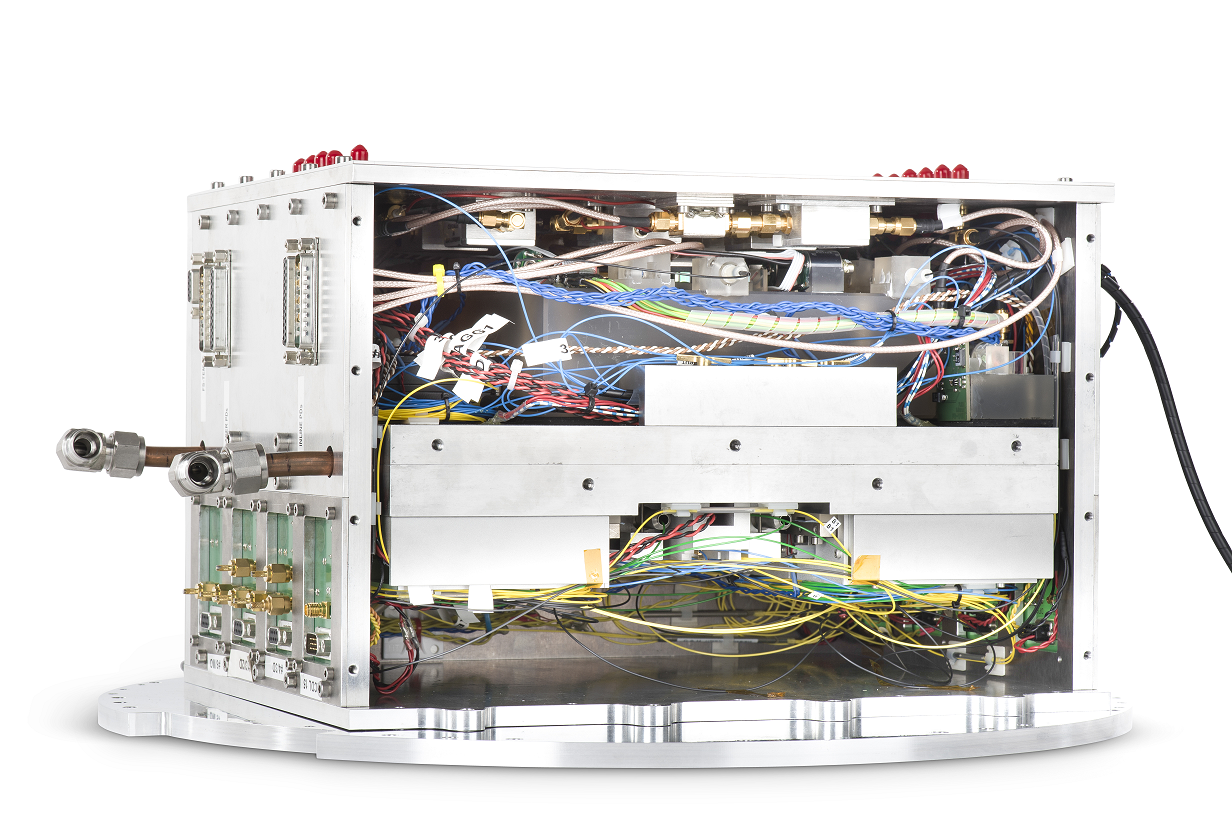}
\caption{The MAIUS laser system with one front cover removed. The science lasers are mounted on the bottom side of the heat sink and the Zerodur benches for switching and frequency stabilization are mounted on top. The dimensions are \SI{340}{mm}$\times$\SI{274}{mm}$\times$\SI{227}{mm} and the total mass is \SI{27}{kg}.}
\label{fig:Maius_LS}
\end{figure}

The laser system delivers \SI{100}{mW} and \SI{10}{mW} for cooling and repumping, respectively, to the 2D+ MOT and \SI{105}{mW} and \SI{10}{mW} for the 3D MOT. The power for each Bragg interferometry beam is higher than \SI{8}{mW}. Thanks to the robust design of the laser system assembly no power stabilization techniques are required.

\subsection{Assembled laser system}

The subsystems presented above are integrated into one aluminum housing providing electrical interfaces to the laser electronics, as well as 11 optical fibers connected to the physics package. The main mechanical structure is a water cooled heat sink, to which all laser modules are mounted on the bottom side. The top side houses the Zerodur switching board, the spectroscopy module, the fiber splitter modules for distribution and frequency stabilization together with electronic parts for processing the microwave beat signals. 

A picture of the assembled flight model is shown in Figure \ref{fig:Maius_LS}. Its dimensions are \SI{340}{mm}$\times$\SI{274}{mm}$\times$\SI{227}{mm} with a total mass of \SI{27}{kg}. All instruments of the MAIUS payload are mounted to standardized platforms, each connected to the rocket hull by six brackets. These brackets contain a passive vibration isolation, which has been developed for MAIUS \cite{Grosse2014}.

The thermal design of the heat-sink results in a temperature increase by less than \SI{2.5}{K} during the flight caused by the heat produced from the laser modules and radiation from the rocket hull, which is compatible with the temperature control system for the laser system. In laboratory environment the convection cooling is sufficient for the laser system to be operated without water cooling.

\section{Qualification and sub system performance tests}
\label{sec:qualification}

The fully assembled laser system was subjected to a vibration acceptance test to simulate the load during the launch. All components beforehand successfully passed the qualification loads of \SI{8.1}{g_{RMS}} hard-mounted or even \SI{29}{g_{RMS}} in the case of the micro-integrated diode laser modules \cite{Luvsandamdin2014}. Second, a successfully completed rocket mission with a laser sub-system including the MAIUS reference laser and the spectroscopy cell assembly in April 2015 was performed. These tests and their results are presented below. In addition, we operated another reference laser based on modulation transfer spectroscopy (MTS) of rubidium aboard the TEXUS 53 mission in January 2016.
\subsection{Vibration test}

For the vibration test the complete, packaged laser system as depicted in Figure \ref{fig:Maius_LS}, was mounted into a test rocket hull segment via shock mounts. The vibration load was set to the acceptance level of \SI{5.4}{g_{RMS}}. The standard test procedure starts and ends with a resonance scan (\SI{5}{} - \SI{2000}{Hz}, \SI{0.25}{g}, \SI{2}{oct/min}) for each axis. In between, the random vibration test lasts for \SI{60}{s} with the specifications provided by the operator corresponding to \SI{5.4}{g_{RMS}}. These tests were conducted with the 2D+ science laser in operation. No degradation or malfunction of the optomechanical assembly was observed in the resonance scans.

To demonstrate the high intrinsic stability of the laser system even further the output power of the 2D+ MOT beam path, which is the longest path in the system, was monitored during the entire test cycle. Neither the output power of the MOPA module running at the working point around \SI{780.24}{nm} nor the fiber output power showed any degradation during the vibration test. After the vibration test, the output power of the remaining paths were compared to their reference measurements done before transport to the shaker facility and showed no decrease within the measurement precision. Since the qualification of the laser system was completed, it has been operated as part of the MAIUS payload for more than 2 years.

\begin{figure}[h]
\centering
\includegraphics[width=0.7\linewidth]{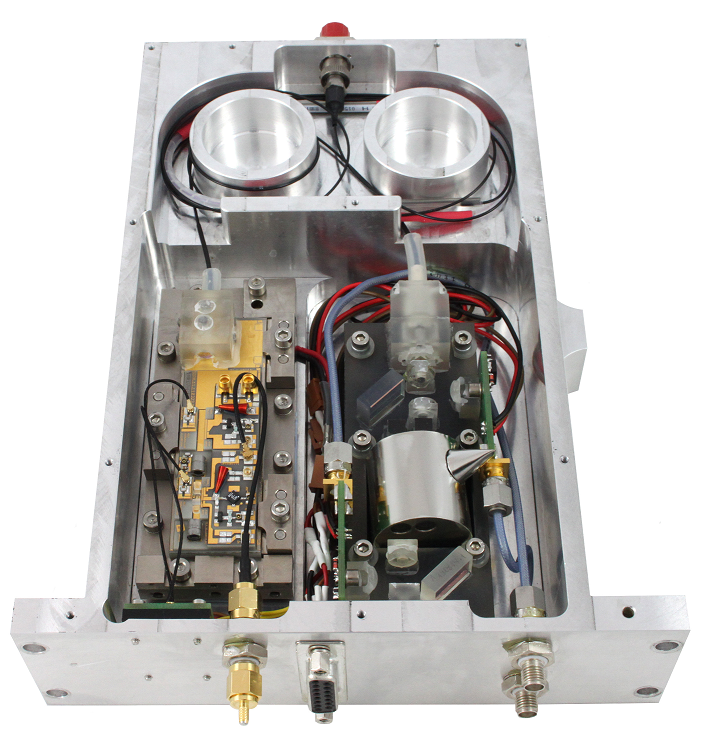}
\caption{The rubidium spectroscopy module for the FOKUS payload with open cover. The frequency of a DFB laser (left) is stabilized to the rubidium gas cell on the optical bench (right). Electronic interfaces for module operation are on the front and the fiber interface with the frequency comb is located on the rear side. The dimensions are \SI{215}{mm}$\times$\SI{170}{mm}$\times$\SI{50}{mm} and the total mass is \SI{2.1}{kg}.}
\label{fig:FOKUS}
\end{figure}

\subsection{Precision Rubidium spectroscopy in space}
\label{subsec:FOKUS}

To demonstrate a high technology readiness level (TRL) for the developed laser system technology, we successfully operated a subsystem assembly on a separate VSB-30 sounding rocket mission. A DFB laser and a spectroscopy bench identical to the ones used in the MAIUS laser system where built and launched as part of the FOKUS payload on the TEXUS 51 sounding rocket mission in April 2015 that was carried out at the Esrange Space Center, Sweden.  FOKUS was built by Menlo Systems and includes a fiber based frequency comb. A picture of this piggyback payload is shown in Figure\,\ref{fig:FOKUS}.

Here, light from the DFB laser diode is fiber coupled and split with a 10:90 ratio using a fiber splitter. The larger part is used for the beat measurement with a frequency comb while the remaining laser light is guided to the Zerodur optical bench including a rubidium gas cell for stabilization of the laser frequency. The rubidium  cell is placed inside a $\mu$-metal shield to reduce frequency shifts.

\begin{figure}[h]
\centering
\includegraphics[width=0.99\linewidth]{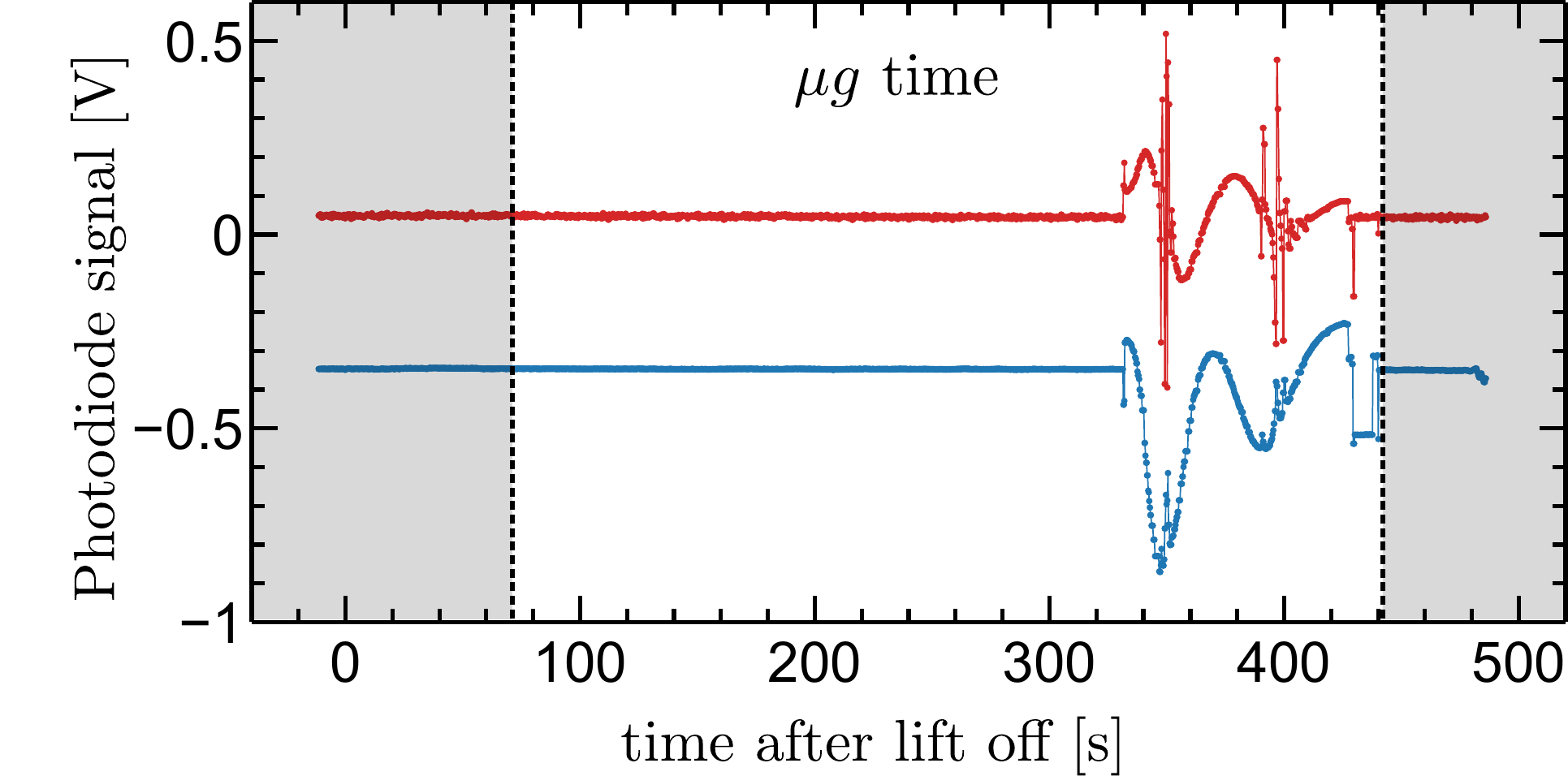}
\caption{The flight data of the frequency stabilized DFB laser on board the FOKUS mission. The error signal (red) and the DC transmission (blue) were recorded with a rate of \SI{3}{Hz}. The unshaded area corresponds to the $\mu g$ time, which begins \SI{70}{s} after lift-off and lasts for almost \SI{350}{s}. The laser remained frequency stabilized during boost and re-entry phase.}
\label{fig:FOKUS_in_space}
\end{figure}

The laser frequency was stabilized before lift-off onto the $\ket{F = 2} \rightarrow \ket{F' = 2/3}$ crossover transition of $^{87}$Rb using an automated locking scheme and the error signal of the frequency lock together with the DC transmission signal were monitored with a rate of \SI{3}{Hz} during the flight. The flight data is shown in Figure \ref{fig:FOKUS_in_space}. 

The laser frequency remained stabilized during the complete boost phase despite the high peak thrust acceleration of \SI{8.1}{g} and more than \SI{12}{g} during each motor separation \cite{OHB}. At \SI{330}{s} after lift-off the laser was intentionally scanned over the rubidium D2 line revealing its recognizable spectroscopic features and was relocked afterward. Full control over the laser's frequency was thereby demonstrated. 

After the payload recovery the reference module was switched on and locked again to demonstrate the functionality of the system. A report about the FOKUS experiment including the full description of the payload and the analysis of the frequency measurement between the comb and the DFB laser in context of a Local Position Invariance (LPI) pathfinder test in space is in preparation \cite{Lezius2016}. The successful mission results in a TRL 9 for sounding rocket missions \cite{ISO} and shows that the presented technology used for the MAIUS laser system can be operated even in such harsh environments.

\section{Conclusion}
\label{sec:summary}

We presented a laser system based on micro integrated diode lasers for atom interferometry in space during a sounding rocket flight. The system passed all 
required tests for launch operation. Additionally, precision laser spectroscopy was performed during a sounding rocket flight using a simplified laser system. With a flight apogee of \SI{258}{km} this is to our knowledge the first Doppler free laser spectroscopy experiment ever performed in space.

The laser system realized for the MAIUS mission, and in particular the technologies developed for its realization pave the way for potential future deployment of complex laser systems, required for atom interferometry based space geodesy, gravitational wave detection and quantum tests of the equivalence principle. The development of laser systems required for such space borne quantum sensors will also benefit compact and mobile atomic sensors on ground. The technology can easily be adapted to various wavelengths for applications with other atomic species.

\section*{Acknowledgments}

We want to thank the Germany Space Agency (DLR) for their support. Special thanks go to Dr. Rainer Kuhl (DLR) for his enthusiasm, motivation and guidance. We thank Menlo Systems for integrating the rubidium spectroscopy module into the FOKUS payload and operation throughout the joint sounding rocket mission.

This work is supported by the German Space Agency DLR with funds provided by the Federal Ministry for Economic Affairs and Energy under grant numbers DLR 50WM 1133, 1237, 1238 and 1345.



\begin{thebibliography}{10}

\bibitem{Kasevich1991a}
M.~Kasevich and S.~Chu, ``{Atomic interferometry using stimulated Raman
  transitions},'' {\em Phys. Rev. Lett.}, vol.~67, pp.~181--184, jul 1991.

\bibitem{Peters2001}
A.~Peters, K.~Y. Chung, and S.~Chu, ``{High-precision gravity measurements
  using atom interferometry},'' {\em Metrologia}, vol.~38, pp.~25--61, feb
  2001.

\bibitem{McGuirk2002}
J.~McGuirk, G.~Foster, J.~Fixler, M.~Snadden, and M.~Kasevich, ``{Sensitive
  absolute-gravity gradiometry using atom interferometry},'' {\em Phys. Rev.
  A}, vol.~65, feb 2002.

\bibitem{Gustavson2000}
T.~L. Gustavson, A.~Landragin, and M.~A. Kasevich, ``{Rotation sensing with a
  dual atom-interferometer Sagnac gyroscope},'' {\em Class. Quantum Gravity},
  vol.~17, pp.~2385--2398, jun 2000.

\bibitem{Freier2015}
C.~Freier, M.~Hauth, V.~Schkolnik, B.~Leykauf, M.~Schilling, H.~Wziontek, H.-G.
  Scherneck, J.~M{\"{u}}ller, and A.~Peters, ``{Mobile quantum gravity sensor
  with unprecedented stability},'' {\em arXiv:1512.05660}, p.~6, dec 2015.

\bibitem{Merlet2010}
S.~Merlet, Q.~Bodart, N.~Malossi, A.~Landragin, F.~P.~D. Santos, O.~Gitlein,
  and L.~Timmen, ``{Comparison between two mobile absolute gravimeters: optical
  versus atomic interferometers},'' {\em Metrologia}, vol.~47, pp.~L9--L11, aug
  2010.

\bibitem{Gillot2014}
P.~Gillot, O.~Francis, A.~Landragin, F.~{Pereira Dos Santos}, and S.~Merlet,
  ``{Stability comparison of two absolute gravimeters: optical versus atomic
  interferometers},'' {\em Metrologia}, vol.~51, p.~L15, oct 2014.

\bibitem{Fang2016}
B.~Fang, I.~Dutta, P.~Gillot, D.~Savoie, J.~Lautier, B.~Cheng, C.~L.~G. Alzar,
  R.~Geiger, S.~Merlet, F.~P.~D. Santos, and A.~Landragin, ``{Metrology with
  Atom Interferometry: Inertial Sensors from Laboratory to Field
  Applications},'' {\em arXiv:1601.06082}, p.~7, jan 2016.

\bibitem{Peters1999}
A.~Peters, K.~Y. Chung, and S.~Chu, ``{Measurement of gravitational
  acceleration by dropping atoms},'' {\em Nature}, vol.~400, pp.~849--852, aug
  1999.

\bibitem{Muller2010}
H.~M{\"{u}}ller, A.~Peters, and S.~Chu, ``{A precision measurement of the
  gravitational redshift by the interference of matter waves.},'' {\em Nature},
  vol.~463, pp.~926--9, feb 2010.

\bibitem{Schlippert2014}
D.~Schlippert, J.~Hartwig, H.~Albers, L.~Richardson, C.~Schubert, A.~Roura,
  W.~Schleich, W.~Ertmer, and E.~Rasel, ``{Quantum Test of the Universality of
  Free Fall},'' {\em Phys. Rev. Lett.}, vol.~112, p.~203002, may 2014.

\bibitem{Zhou2015}
L.~Zhou, S.~Long, B.~Tang, X.~Chen, F.~Gao, W.~Peng, W.~Duan, J.~Zhong,
  Z.~Xiong, J.~Wang, Y.~Zhang, and M.~Zhan, ``{Test of Equivalence Principle at
  10(-8) Level by a Dual-Species Double-Diffraction Raman Atom
  Interferometer.},'' {\em Phys. Rev. Lett.}, vol.~115, p.~013004, jul 2015.

\bibitem{Muentinga2013}
H.~M\"untinga, H.~Ahlers, M.~Krutzik, A.~Wenzlawski, S.~Arnold, D.~Becker,
  K.~Bongs, H.~Dittus, H.~Duncker, N.~Gaaloul, C.~Gherasim, E.~Giese,
  C.~Grzeschik, T.~W. H\"ansch, O.~Hellmig, W.~Herr, S.~Herrmann, E.~Kajari,
  S.~Kleinert, C.~L\"ammerzahl, W.~Lewoczko-Adamczyk, J.~Malcolm, N.~Meyer,
  R.~Nolte, A.~Peters, M.~Popp, J.~Reichel, A.~Roura, J.~Rudolph,
  M.~Schiemangk, M.~Schneider, S.~T. Seidel, K.~Sengstock, V.~Tamma,
  T.~Valenzuela, A.~Vogel, R.~Walser, T.~Wendrich, P.~Windpassinger, W.~Zeller,
  T.~van Zoest, W.~Ertmer, W.~P. Schleich, and E.~M. Rasel, ``Interferometry
  with bose-einstein condensates in microgravity,'' {\em Phys. Rev. Lett.},
  vol.~110, p.~093602, Feb 2013.

\bibitem{Dickerson2013}
S.~M. Dickerson, J.~M. Hogan, A.~Sugarbaker, D.~M.~S. Johnson, and M.~A.
  Kasevich, ``{Multiaxis Inertial Sensing with Long-Time Point Source Atom
  Interferometry},'' {\em Phys. Rev. Lett.}, vol.~111, p.~083001, aug 2013.

\bibitem{Hartwig2015}
J.~Hartwig, S.~Abend, C.~Schubert, D.~Schlippert, H.~Ahlers, K.~Posso-Trujillo,
  N.~Gaaloul, W.~Ertmer, and E.~M. Rasel, ``{Testing the universality of free
  fall with rubidium and ytterbium in a very large baseline atom
  interferometer},'' {\em New J. Phys.}, vol.~17, p.~035011, mar 2015.

\bibitem{Aguilera2014}
D.~N. Aguilera, H.~Ahlers, B.~Battelier, A.~Bawamia, A.~Bertoldi,
  R.~Bondarescu, K.~Bongs, P.~Bouyer, C.~Braxmaier, L.~Cacciapuoti,
  C.~Chaloner, M.~Chwalla, W.~Ertmer, M.~Franz, N.~Gaaloul, M.~Gehler,
  D.~Gerardi, L.~Gesa, N.~G{\"{u}}rlebeck, J.~Hartwig, M.~Hauth, O.~Hellmig,
  W.~Herr, S.~Herrmann, A.~Heske, A.~Hinton, P.~Ireland, P.~Jetzer, U.~Johann,
  M.~Krutzik, A.~Kubelka, C.~L{\"{a}}mmerzahl, A.~Landragin, I.~Lloro,
  D.~Massonnet, I.~Mateos, A.~Milke, M.~Nofrarias, M.~Oswald, A.~Peters,
  K.~Posso-Trujillo, E.~Rasel, E.~Rocco, A.~Roura, J.~Rudolph, W.~Schleich,
  C.~Schubert, T.~Schuldt, S.~Seidel, K.~Sengstock, C.~F. Sopuerta,
  F.~Sorrentino, D.~Summers, G.~M. Tino, C.~Trenkel, N.~Uzunoglu, W.~von
  Klitzing, R.~Walser, T.~Wendrich, A.~Wenzlawski, P.~We{\ss}els, A.~Wicht,
  E.~Wille, M.~Williams, P.~Windpassinger, and N.~Zahzam, ``{STE-QUEST test of
  the universality of free fall using cold atom interferometry},'' {\em Class.
  Quantum Gravity}, vol.~31, p.~115010, jun 2014.

\bibitem{Graham2013}
P.~W. Graham, J.~M. Hogan, M.~A. Kasevich, and S.~Rajendran, ``New method for
  gravitational wave detection with atomic sensors,'' {\em Phys. Rev. Lett.},
  vol.~110, p.~171102, Apr 2013.

\bibitem{Yu2006}
N.~Yu, J.~Kohel, J.~Kellogg, and L.~Maleki, ``{Development of an
  atom-interferometer gravity gradiometer for gravity measurement from
  space},'' {\em Appl. Phys. B}, vol.~84, pp.~647--652, jul 2006.

\bibitem{Carraz2014}
O.~Carraz, C.~Siemes, L.~Massotti, R.~Haagmans, and P.~Silvestrin, ``{A
  Spaceborne Gravity Gradiometer Concept Based on Cold Atom Interferometers for
  Measuring Earth’s Gravity Field},'' {\em Microgravity Sci. Technol.},
  vol.~26, pp.~139--145, oct 2014.

\bibitem{Hogan2011}
J.~M. Hogan, D.~M.~S. Johnson, S.~Dickerson, T.~Kovachy, A.~Sugarbaker, S.-w.
  Chiow, P.~W. Graham, M.~A. Kasevich, B.~Saif, S.~Rajendran, P.~Bouyer, B.~D.
  Seery, L.~Feinberg, and R.~Keski-Kuha, ``{An atomic gravitational wave
  interferometric sensor in low earth orbit (AGIS-LEO)},'' {\em General
  Relativity and Gravitation}, vol.~43, pp.~1953--2009, May 2011.

\bibitem{Geiger2011}
R.~Geiger, V.~M{\'{e}}noret, G.~Stern, N.~Zahzam, P.~Cheinet, B.~Battelier,
  A.~Villing, F.~Moron, M.~Lours, Y.~Bidel, A.~Bresson, A.~Landragin, and
  P.~Bouyer, ``{Detecting inertial effects with airborne matter-wave
  interferometry},'' {\em Nat. Commun.}, vol.~2, p.~474, sep 2011.

\bibitem{Theron2014}
F.~Theron, O.~Carraz, G.~Renon, N.~Zahzam, Y.~Bidel, M.~Cadoret, and
  A.~Bresson, ``{Narrow linewidth single laser source system for onboard atom
  interferometry},'' {\em Appl. Phys. B}, vol.~118, pp.~1--5, Dec. 2014.

\bibitem{Leveque2014}
T.~L\'{e}v\`{e}que, L.~Antoni-Micollier, B.~Faure, and J.~Berthon, ``{A laser
  setup for rubidium cooling dedicated to space applications},'' {\em Appl.
  Phys. B}, vol.~116, pp.~997--1004, Feb. 2014.

\bibitem{Leveque2015}
T.~L\'{e}v\`{e}que, B.~Faure, F.~X. Esnault, C.~Delaroche, D.~Massonnet,
  O.~Grosjean, F.~Buffe, P.~Torresi, T.~Bomer, A.~Pichon, P.~B\'{e}raud, J.~P.
  Lelay, S.~Thomin, and P.~Laurent, ``{PHARAO laser source flight model: design
  and performances.},'' {\em Rev. Sci. Instrum.}, vol.~86, p.~033104, Mar.
  2015.

\bibitem{Grosse2014}
J.~Grosse, S.~Seidel, M.~Krutzik, M.~Scharringhausen, and T.~van Zoest,
  ``{Thermal and mechanical design of the MAIUS atom interferometer sounding
  rocket payload},'' in {\em AIAA SPACE 2014 Conference and Exposition, SPACE
  Conferences and Exposition}, ESA Communications, okt 2014.

\bibitem{dlr74326}
A.~Garcia, S.~S.~C. Yamanaka, A.~N. Barbosa, F.~C.~P. Bizarria, W.~Jung, and
  F.~Scheuerpflug, ``{VSB-30 sounding rocket: history of flight performance},''
  {\em J. Aerosp. Technol. Manag.}, vol.~Vol.3, pp.~325--330, sep 2011.

\bibitem{Rudolph2015}
J.~Rudolph, W.~Herr, C.~Grzeschik, T.~Sternke, A.~Grote, M.~Popp, D.~Becker,
  H.~M{\"{u}}ntinga, H.~Ahlers, A.~Peters, C.~L{\"{a}}mmerzahl, K.~Sengstock,
  N.~Gaaloul, W.~Ertmer, and E.~M. Rasel, ``{A high-flux BEC source for mobile
  atom interferometers},'' {\em New J. Phys.}, vol.~17, p.~065001, jun 2015.

\bibitem{Muntinga2013}
H.~M{\"{u}}ntinga, H.~Ahlers, M.~Krutzik, A.~Wenzlawski, S.~Arnold, D.~Becker,
  K.~Bongs, H.~Dittus, H.~Duncker, N.~Gaaloul, C.~Gherasim, E.~Giese,
  C.~Grzeschik, T.~W. H{\"{a}}nsch, O.~Hellmig, W.~Herr, S.~Herrmann,
  E.~Kajari, S.~Kleinert, C.~L{\"{a}}mmerzahl, W.~Lewoczko-Adamczyk,
  J.~Malcolm, N.~Meyer, R.~Nolte, A.~Peters, M.~Popp, J.~Reichel, A.~Roura,
  J.~Rudolph, M.~Schiemangk, M.~Schneider, S.~T. Seidel, K.~Sengstock,
  V.~Tamma, T.~Valenzuela, A.~Vogel, R.~Walser, T.~Wendrich, P.~Windpassinger,
  W.~Zeller, T.~van Zoest, W.~Ertmer, W.~P. Schleich, and E.~M. Rasel,
  ``{Interferometry with Bose-Einstein condensates in microgravity.},'' {\em
  Phys. Rev. Lett.}, vol.~110, p.~093602, mar 2013.

\bibitem{Stamminger2015}
A.~Stamminger, J.~Ettl, J.~Grosse, M.~H{\"{o}}rschgen-Eggers, F.~Jung,
  A.~Kallenbach, G.~Raith, W.~Saedtler, S.~Seidel, J.~Turner, and M.~Wittkamp,
  ``{MAIUS-1 - Vehicle, Subsystems Design and Mission Operations},'' in {\em
  Proc. 22nd ESA Symp. Eur. Rocket Balloon Program. Relat. Res.}, vol.~SP-730,
  pp.~183--190, ESA Communications, sep 2015.

\bibitem{Luvsandamdin2014}
E.~Luvsandamdin, C.~K\"{u}rbis, M.~Schiemangk, A.~Sahm, A.~Wicht, A.~Peters,
  G.~Erbert, and G.~Tr\"{a}nkle, ``{Micro-integrated extended cavity diode
  lasers for precision potassium spectroscopy in space.},'' {\em Optics
  express}, vol.~22, pp.~7790--8, Apr. 2014.

\bibitem{Duncker2014}
H.~Duncker, O.~Hellmig, A.~Wenzlawski, A.~Grote, A.~J. Rafipoor, M.~Rafipoor,
  K.~Sengstock, and P.~Windpassinger, ``{Ultrastable, Zerodur-based optical
  benches for quantum gas experiments.},'' {\em Applied optics}, vol.~53,
  pp.~4468--74, July 2014.

\bibitem{Schiemangk2015}
M.~Schiemangk, K.~Lampmann, A.~Dinkelaker, A.~Kohfeldt, M.~Krutzik,
  C.~K\"{u}rbis, A.~Sahm, S.~Spie\ss~berger, A.~Wicht, G.~Erbert,
  G.~Tr\"{a}nkle, and A.~Peters, ``{High-power, micro-integrated diode laser
  modules at 767 and 780 nm for portable quantum gas experiments.},'' {\em
  Applied optics}, vol.~54, pp.~5332--8, June 2015.

\bibitem{Altin2013}
P.~A. Altin, M.~T. Johnsson, V.~Negnevitsky, G.~R. Dennis, R.~P. Anderson,
  J.~E. Debs, S.~S. Szigeti, K.~S. Hardman, S.~Bennetts, G.~D. McDonald, L.~D.
  Turner, J.~D. Close, and N.~P. Robins, ``{Precision atomic gravimeter based
  on Bragg diffraction},'' {\em New J. Phys.}, vol.~15, p.~023009, feb 2013.

\bibitem{Hansel2001}
W.~H{\"{a}}nsel, P.~Hommelhoff, T.~W. H{\"{a}}nsch, and J.~Reichel,
  ``{Bose-Einstein condensation on a microelectronic chip.},'' {\em Nature},
  vol.~413, pp.~498--501, oct 2001.

\bibitem{OHB}
O.~SE, ``{Quicklook Texus 51}.'' Data provided by OHB SE, 2015.

\bibitem{Lezius2016}
M.~Lezius {\em et~al.}, ``{Space-born Frequency Comb Metrology},'' {\em in
  preparation}, 2016.

\bibitem{ISO}
I.~16290:2013, ``{Space systems -- Definition of the Technology Readiness
  Levels (TRLs) and their criteria of assessment}.'' ISO norm, 2013.

\end{thebibliography}
%

\end{document}